\documentclass[%
 reprint,
 amsmath,amssymb,
 aps,
 prl,
 longbibliography
]{revtex4-2}

\usepackage{graphicx}
\usepackage{dcolumn}
\usepackage{bm}
\usepackage{cleveref}
\usepackage{xcolor}
\usepackage{soul}
\newcommand{\fermi}{\text{F}}
\renewcommand{\Im}{\text{Im}}
\begin{document}

\title{Dynamical Josephson Effect Between a Singlet and a Triplet Superconductor}

\author{Morten Amundsen}
 \email{morten.amundsen@ntnu.no}
\affiliation{Center for Quantum Spintronics, Department of Physics,\\
Norwegian University of Science and Technology, NO-7491 Trondheim, Norway
}%
\author{Niladri Banerjee}%
\affiliation{Department of Physics, Blackett Laboratory,
Imperial College London, London SW7 2AZ, United Kingdom
}%
\author{Igor \v{Z}uti\'c}
\affiliation{Department of Physics, University at Buffalo, 
State University of New York, Buffalo, New York 14260, USA
}%

\begin{abstract}
Phase-sensitive Josephson effect has long been central to identifying unconventional pairing symmetries in superconductors. Although the selection rules governing Josephson junctions (JJs) are generally determined by the symmetries of the constituent superconductors, we demonstrate that this paradigm is modified in the dynamic regime. By modeling a JJ 
where spin-singlet and spin-triplet superconductors are separated by a two-dimensional electron gas, we show that a time-dependent gate voltage qualitatively changes the underlying selection rules. This modification arises as a consequence of the gate-controlled spin-orbit coupling. 
A harmonic modulation of the gate voltage generates an oscillatory $\cos \phi$ Josephson component which vanishes in the static limit. 
The resulting charge current contains both dissipationless and dissipative components, with the latter strongly suppressed at low temperatures. 
This dynamical Josephson effect could transform the use of JJs in qubits, as sources of spin-triplet currents, and as platforms for proximity effects.
\end{abstract} 

\maketitle

In Josephson effect (JE) a dissipationless charge current between two superconductors is driven across a nonsuperconducting material by the phase difference of the superconducting order parameters~\cite{josephson1962,Tafuri:2019}. These Josephson junctions (JJs) based on conventional spin-singlet Bardeen-Cooper-Schrieffer (BCS) superconductors are well understood and support macroscopic quantum 
tunneling~\cite{Voss1981:PRL,Devoret1985:PRL} which are 
key to advanced quantum computers~\cite{Krantz2019:APR,Kim2025:X}. However, in JJs with exotic materials or strong spin-orbit coupling (SOC)~\cite{zutic2004,Gorkov2001:PRL} the inherent behavior is much richer, from dissipationless spin currents and nonreciprocal response for 
superconducting spintronics~\cite{linder2015,amundsen2024b,Eschrig2011:PT,Martinez2020:PRA,nadeem2023}, to topological superconductivity~\cite{Dartiailh2021:PRL,Ren2019:N,Fornieri2019:N} for fault-tolerant quantum computing~\cite{DasSarma2015:NPJQI,Flensberg2021:NRM,Zhou2022:NC}.

Many of these interesting manifestations in JJs are associated with the presence of spin-triplet superconductivity, either induced by proximity effects~\cite{Zutic2019:MT,buzdin2005,amundsen2024b} or intrinsic~\cite{ott1983,stewart1984,saxena2000,ran2019,jiao2020,kinjo2023,yoon2024,li2025,levitin2025}. Interestingly, in a singlet-triplet JJ between BCS and triplet superconductors, the mismatch in the spin symmetries of the order parameters suggests no JE to the lowest order in the tunneling~\cite{pals1977}. As a consequence, two such JJs forming a SQUID have a $\pi/2$ phase shift compared to all-singlet JJs~\cite{geshkenbein1987}, suggesting SQUIDs for probing triplet superconductors~\cite{geshkenbein1987,nelson2004,zutic2005,xu2024,yamaki2025}.   

In JE there is an interplay between its DC and AC manifestations: The phase difference, $\phi$, drives the supercurrent, while a finite voltage drives the time evolution of $\phi(t)$ and an alternating supercurrent. Time-dependent perturbations can also change the usual frequency-dependence of the superconducting response and give rise to odd-frequency correlations~\cite{linder2019,triola2016,triola2017,cayao2021}. In this work, we reveal unexplored phenomena in the dynamical response of singlet-triplet JJs which offer a probe of unconventional superconductivity as well a platform for emerging applications. As shown in Fig.~1, we consider JJs where the time-dependent gate voltage, $V_G(t)$, dynamically changes the Rashba SOC strength, $\alpha(t)$. Remarkably, even in the absence of other spin-dependent effects or applied magnetic field, ${\bm B}$, when the difference of the spin symmetries precludes the static JE, we find that singlet-triplet JJs supports novel current-phase relation (CPR) $\propto \cos \phi$, in contrast to the all-singlet CPR $\propto \sin \phi$. 

 We consider $\alpha(t)$ in a 2DEG which is supported by the experiments in Al/InAs-based JJs at $B\neq0$ 
where changing $V_G$ transforms CPR $\propto \sin(\phi+\phi_0)$ and tunes the anomalous phase $\phi_0 \propto \alpha$~\cite{Mayer2020:NC}. 
Furthermore, 
at $B\neq0$, $V_G(t)$ which controls $\alpha(t)$, could turn {\em on} and {\em off} JJs and even no applied bias and create jumps in $\phi$~\cite{monroe2022,monroe2024}. Generally, $V_G$  changes 
$\alpha$ and the chemical potential, $\mu$, but the presence of the back gate can fix 
$\mu$~\cite{Papadakis1999:S,VanTuan2019:PRB,Zhang2023:N}. 
       
\begin{figure}[t!]
\includegraphics{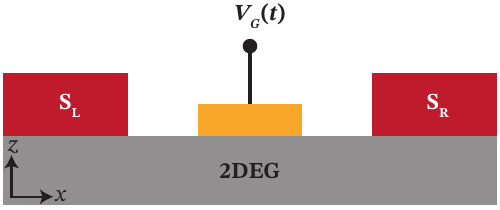}
\caption{Schematic of the studied planar Josephson junction including singlet BCS (S$_\text{L}$) and triplet (S$_\text{R}$) superconductors, connected by the two-dimensional electron gas (2DEG) where the time-dependent gate voltage, $V_G(t)$, gives rise to a time-dependent spin-orbit coupling.}
\label{fig:model}
\end{figure} 

The mechanism of our predicted dynamical JE can be understood even at $B=0$ by assuming a simple form of SOC, $\alpha(t) k_y\sigma_z$, where $k_y$ is the
wave vector along the $y$ axis and $\sigma_z$ is the spin along $z$. This corresponds to an emergent SU(2) vector potential $\bm{A} = \alpha(t)\sigma_z \hat{y}$. We assume $S_R$ to be the $p_y$-wave triplet superconductor, with the $d$ vector pointing along $z$, so that spins are conserved. $\bm{A}(t)$ implies an emergent electric field $\bm{E} = -\partial_t \bm{A}$. Hence, for any trajectory with direction $\hat{n}$ such that $\hat{n}\cdot\hat{y}\neq0$, there is a finite spin-voltage bias between the $S_L$ and $S_R$.
This means that spin-up $\uparrow$ (spin-down $\downarrow$) electrons experience bias $V>0$ ($V<0$). 
In a singlet $S_L$, the order parameter is $\Delta^{(s)} \propto \langle a_\downarrow a_\uparrow\rangle - \langle a_\uparrow a_\downarrow\rangle$,
with $a_{\uparrow,\downarrow}$ annihilation operators,
where the first term is a correlation between a hole$\,\downarrow$ and a electron$\,\uparrow$, which only involves 
$\uparrow$, particles. Similarly, the second term involves only 
$\downarrow$ particles, implying a $\pi$ phase difference between the two spin species. In $S_R$ the triplet $\Delta^{(t)} \propto \langle a_\downarrow a_\uparrow\rangle + \langle a_\uparrow a_\downarrow\rangle$ does not feature such a phase difference. This means that in singlet-triplet JJs, supercurrents of opposite spin flow in opposite directions. This sign difference is compensated by the spin voltage induced by the driven SOC, producing an effective AC JE, $d\phi/dt\propto$ applied voltage~\cite{Tafuri:2019}.  
We model the system in Fig.~\ref{fig:model} 
by a Hamiltonian
\begin{align}
H = H_{L} + H_{R} + H_T, 
\label{eq:hamiltonian}
\end{align}
as a sum of the left, right, and the tunneling part with
\begin{align}
H_{j} &= \sum_{ks} \xi_k a_{j,ks}^\dagger a_{j,ks} + \sum_{kss'}\Delta_{j,ss'}a^\dagger_{j,ks}a^{\dagger}_{j,-ks'} + h.c., 
\label{eq:Hj}
\end{align}
where 
$\xi_k = \hbar^2k^2/2m^* - \mu$, with effective mass $m^*$,
$\mu = \hbar^2 k_\fermi^2/2m^*$, and $k_\fermi$ the Fermi wave vector, 
while $a_{j,ks}^\dagger$ creates a particle in 
$S_j$, with $\Delta_j$, $j\in\left\{L,R\right\}$, and $s \in \left\{\uparrow, \downarrow \right\}$.
$\Delta_L$ is a $s$-wave singlet, uniform in ${\bm k}$ space, 
and $\Delta_R = \Delta_{R,0}\sin\theta$, with $\sin\theta = k_y/k_\fermi$ is the chosen
$p$-wave triplet. 
We assume that the two interfaces between the three materials are $\|\hat{y}$, so that $k_y$ is conserved.

Tunneling between the $S_L$ and $S_R$ is given by $H_T$
\begin{align}
H_{T} = \sum_{kqs} \left(T_{kqs}a^\dagger_{R,ks}a_{L,qs} + T^*_{kqs}a^\dagger_{L,qs}a_{R,ks}\right),
\end{align}
with $T_{kqs}$ indicating tunneling between $\bm{q}$ in $S_L$ and $\bm{k}$ in $S_R$ for spin $s$.
The current flowing into (or out of) $S_L$ is defined as the rate of change in the charge density of this material, $e$ is the electron charge
\begin{align}
I_L(t) = -e\left\langle \dot{n}_L\right\rangle = \frac{ie}{\hbar}\sum_{ks}\left\langle\left[a^\dagger_{L,ks} a_{L,ks},H_T\right]\right\rangle,
\end{align} 
ignoring effects which do not involve tunneling. To understand the underlying physics we employ a simple SOC model 
$\propto \sigma_z k_y$ in a 2DEG described by a spin and direction-dependent phase accumulated as particles tunnel from one side to the other. In our JJs, this yields 
\begin{align}
T_{kqs}(t) = T_{kq} e^{ius\alpha(t)\sin\theta},
\end{align} 
with $u = \text{sgn}(k_x+q_x)$ indicating the particle traveling left to right $(u = 1)$ or right to left $(u = -1)$, while $s=1$ $(-1)$ for $\uparrow$ ($\downarrow$). We use the common approximation that $|T_{kq}|$ is independent of $k$ and $q$~\cite{Mahan:2000}, and define 
$|T_{kq}|^2 = \tau / A^2\nu_0^2$, with $\tau$ the transparency, $A$ the junction area, and $\nu_0$ the density of states of the 2DEG, evaluated at $\mu$. This does not fully capture the JE magnitude, 
but is sufficient to predict the CPR.
\begin{figure}[t!]
\includegraphics{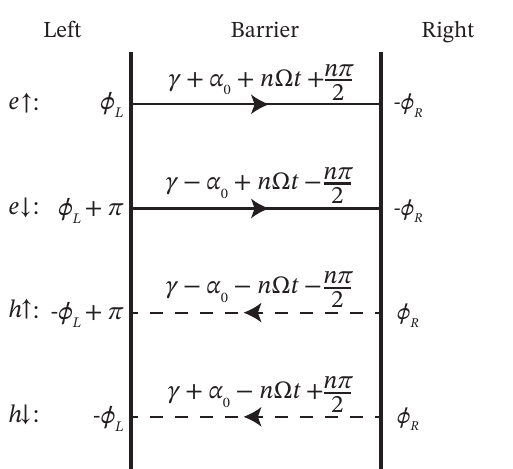}
\caption{
Schematic of the phase accumulated by electrons$\,{\uparrow,\downarrow}$ tunneling from the singlet to the triplet superconductor, and holes$\,{\uparrow,\downarrow}$ tunneling in the opposite direction.}
\label{fig:phase}
\end{figure}
We will assume a time-dependent SOC modulation $\alpha(t) = \alpha_0 + \alpha_1\cos\Omega t$. By the Jacobi-Anger expansion~\cite{abramowitz2014} we find
\begin{align}
T_{kqs}(t) &= \sum_{n=-\infty}^{\infty} T_{kqs}^{(n)} e^{in\Omega t},
\label{eq:tamp}\\
T_{kqs}^{(n)} &= i^n T_{kqs}^0 J_n(su\alpha_1\sin\theta),
\end{align}
where $J_n(x)$ is the Bessel function. The time-dependent modulation enters the $H_T$ in a similar way as microwave radiation~\cite{Tafuri:2019}. However, the present system is
more complicated: The singlet-triplet ($S_L$-$S_R$) coupling enforces specific selection rules for permissible frequency modes $n$. To understand why, recall that the normal state current through a tunneling barrier $ \propto |T_{kqs}|^2$. For JE, $I$ is  
transmitted by Andreev bound states~\cite{Tafuri:2019}, and this electron-hole coupling leads instead to $I \propto \Im\left(T_{kqs} T_{\bar{k}\bar{q}\bar{s}}\right)$, with $\bar{x} = -x$. The phase accumulated by the tunneling particle is important. We can infer the CPR directly by studying tunneling processes in Fig.~\ref{fig:phase}. For the static limit, $\Omega = 0$,  an electron$\,\uparrow$ tunneling through the barrier will accumulate a spin-independent phase $\gamma \propto\;$ the barrier width, and a spin-dependent phase $\alpha$ due to the static SOC, in addition to $\phi_L-\phi_R$. With $T_{kq\uparrow} = |T_{kq}|e^{i\psi_{kqs}}$,  we get 
$\psi_{kq\uparrow} = \gamma + \alpha_0 + (\phi_L - \phi_R)/2$.  If we reverse the tunneling direction: $\bm{k}\to-\bm{k}$ and $\bm{q}\to-\bm{q}$, $\gamma\propto\text{sgn}(k_x+q_x)$ reverses its sign, 
unlike $\alpha_0\propto\text{sgn}[k_y(k_x+q_x)]$. This means that a hole$\,\downarrow$  will accumulate a phase $\psi_{\bar{k}\bar{q}\downarrow} = -\gamma-\alpha_0 -(\phi_L-\phi_R)/2$, and that $\Im\left(T_{kq\uparrow}T_{\bar{k}\bar{q}\downarrow}\right) \propto \sin(\phi_L-\phi_R)$, independent of both $\gamma$ and $\alpha_0$. 

We can apply the same procedure to electrons$\,\downarrow$, with the only modification that, due to the spin-singlet structure of $S_L$, we get an additional phase $\pi$ relative to the $\uparrow$ case. Therefore, $\Im\left(T_{kq\downarrow}T_{\bar{k}\bar{q}\uparrow}\right)\propto -\sin(\phi_L-\phi_R)$. Summing over spins reveals that there are no net charge currents in this system, even in the presence of SOC of matching spin and momentum symmetries, while there are spin currents, that implies {\em pure} spin currents~\cite{zutic2004}.

With $\Omega\neq0$ the system becomes more complicated. We get a spin-independent phase contribution, $n\Omega t$, which is a direct consequence of $H=H(t)$. The spin structure of this perturbation introduces an additional phase factor $sn\pi/2$ 
for spin $s$, from Eq.~(\ref{eq:tamp}). Adding up all the phases from Fig.~\ref{fig:phase} gives 
the current contribution
\begin{align}
&I_{nm} = \Im\left(T^{(n)}_{kq\uparrow}T^{(m)}_{\bar{k}\bar{q}\downarrow} + T^{(n)}_{kq\downarrow}T^{(m)}_{\bar{k}\bar{q}\uparrow}\right) \propto\\  \nonumber
& \sin\frac{(n-m)\pi}{2}\cos\left[\phi + (n+m)\Omega t\right], 
\end{align}
with the $I = \sum_{mn} I_{mn}$. Thus, we see that JE emerges only when $n-m = 2k+1$, with $k$ an integer. From the orthogonality of $J_n$, We note that $I\rightarrow 0$ for $\Omega\to 0$. 
We can extract more information from this expression. Since we sum over all positive and negative $n$ and $m$, we may write $I_{nm} = \left(I_{nm} + I_{\bar{n}\bar{m}}\right)/2$ and, using that $J_{\bar{n}}(x) = (-1)^n J_n(x)$, we find
\begin{align}
I_{nm} \propto \sin\frac{(n-m)\pi}{2}\cos \phi\cos\left[(n+m)\Omega t\right].
\end{align}
We predict a dynamical JE with a cosine CPR, 
oscillating at odd multiples of the driving frequency. This even-in-phase response in the dynamic regime is a direct consequence of opposite spin symmetry of the condensate and the effect disppears in the static limit. By repeating this analysis for a singlet-singlet JJ (see Supplemental Material~\cite{SM}), provides a 
useful comparison
\begin{align}
I^{ss}_{nm} \propto \cos\frac{(n-m)\pi}{2}\sin\phi \cos\left[(n+m)\Omega t\right],
\end{align}
with the selection rule $n-m = 2k$. For $n=m=0$ we recover the
conventional JE and the CPR $\propto \sin \phi$.

Our above analysis offers both a simple way to deduce novel CPRs as well as recover known results~\cite{buzdin2005}. However, it does not provide information about the magnitude of $I$. In addition, the explicit time dependence allows for tunneling between different energies. This gives rise to an additional contribution to the JE not captured in the above analysis, which is a dissipative current, $I_A$. To verify our predictions and obtain these missing pieces we study $H$ in Eq.~(\ref{eq:hamiltonian}) using the Keldysh Green function formalism~\cite{Zagoskin:2014,SM}. For the singlet-triplet JJ, we
find
\begin{widetext}
\begin{align}
\label{eq:JJcurrent}
I(t) &= \frac{\pi|\Delta_L\Delta_{0,R}|\cos\phi}{2 eR_N}\int_{-\pi/2}^{\pi/2}d\theta\;\Gamma^{10}(\theta)\sin\theta\left[\left(I_\omega(\Omega) - I_\omega(0)\right)\cos\Omega t + J_\omega(\Omega)\sin\Omega_0 t   \vphantom{\frac{1}{1}}\right]\\
&\equiv I_S\cos\phi\cos\Omega t + I_A\cos\phi\sin\Omega t,\nonumber
\end{align}
decomposing the current into dissipationless (symmetric) and dissipative (antisymmetric) contributions in $t$, with 
\begin{align*}
I_\omega(\Omega) &=\int d\omega \;\Im\left[\frac{T_+(\omega)}{\sqrt{|\Delta_L|^2-(\omega_++i\delta)^2}\sqrt{|\Delta_R|^2-(\omega_-+i\delta)^2}} + \frac{T_-(\omega)}{\sqrt{|\Delta_L|^2-(\omega_++i\delta)^2}\sqrt{|\Delta_R|^2-(\omega_--i\delta)^2}}\right], \\
J_\omega(\Omega) &=\int d\omega \;\Im\left[\frac{1}{\sqrt{|\Delta_L|^2-(\omega_++i\delta)^2}}\right] \Im\left[\frac{1}{\sqrt{|\Delta_R|^2-(\omega_-+i\delta)^2}}\right]T_-(\omega),
\end{align*}
\end{widetext}
where $\omega_\pm = \omega \pm \Omega/2$, and 
$T_\pm(\omega) = (1/2)\tanh (\beta\omega_+/2)\pm (1/2)\tanh (\beta\omega_-)/2$, for $\beta = 1/k_\text{B}T$, with $T$ the temperature and $k_{\text{B}}$ the Boltzmann constant. Furthermore, $R_N = h/2 e^2\tau$ is the normal-state tunneling resistance. We also defined $\Gamma^{10}(\theta) = J_0(\alpha_1 u \sin\theta)J_1(\alpha_1 u \sin\theta)$~\cite{SM}. At $T=0$, $I_\omega(0)$ yields a form similar to the standard Ambegaokar-Baratoff critical current~\cite{ambegaokar1963}
\begin{align}
I_\omega(0) = \frac{2}{|\Delta_L|} K\left(\sqrt{1-|\Delta_R(\theta)|^2/|\Delta_L|^2}\right),
\end{align}
with $K(x)$ the complete elliptic integral of the first kind.

From Eq.~(\ref{eq:JJcurrent}), we see that a term 
$I_S$ is consistent with the conventional AC JE, maximal when $\phi_L-\phi_R=0$. Unlike in conventional singlet-singlet JJs, this contribution, which oscillates in phase with the driving, is 
dissipationless because the effective bias that generates this current is given by $V(t) \propto \dot{\alpha} \propto \alpha_1\sin\Omega t$, with the averaged dissipated
power $P = \langle V(t)I(t) \rangle= 0$.

For the singlet-singlet JJ, Eq.~(\ref{eq:JJcurrent}) becomes
\begin{align}
I(t) = \left[I^{ss}_0 + I^{ss}_c\cos 2\Omega t + I^{ss}_s \sin 2\Omega t\right]\sin\phi,
\label{eq:sscurrent}
\end{align}
with $I^{ss}_0$, $I^{ss}_S$ and $I^{ss}_A$, explicitly given in ~\cite{SM}. The presence of $I_0^{ss}$ and $I_S^{ss}$, as well as the frequency dependence is again as predicted by our simplified approach from Fig.~\ref{fig:phase}.
What the phenomenological analysis of the tunneling phases did not reveal is the dissipative contribution $I_A$. At $T=0$,  
$I_A=0$ for $\hbar\,\Omega < \Delta$. 
 
\begin{figure}[t!]
\includegraphics{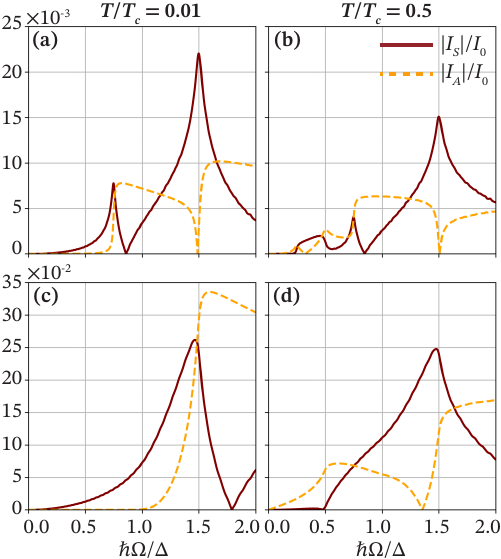}
\caption{The dissipationless (dissipative) $I_S$ ($I_A$) current amplitude 
as a function of the
driving frequency $\Omega$, normalized to 
$I_0 = \pi |\Delta_{0,R}|/2eR_N$. 
(a), (b) for singlet-singlet and (c), (d) singlet-triplet JJs.}
\label{fig:current}
\end{figure}

Next, we numerically evaluate Eqs.~(\ref{eq:JJcurrent}), (\ref{eq:sscurrent}) for a range of  $\Omega$. We interpret the quantity $\delta$ as a Dynes parameter~\cite{dynes1978}, incorporating the effects of inelastic scattering, and choose $\delta = 0.01$. We let $\Delta_L = \Delta$ as the energy scale of our system
and choose the amplitude of 
the order parameter in $S_R$ to be $\Delta_{R,0} = 0.5\Delta$, and the SOC strength $\alpha_1 k_F / \Delta = 0.1$,
consistent with the SOC in the 
JJs based on the gated InAs 2DEG~\cite{Mayer2020:NC,Dartiailh2021:PRL}. 

For singlet-singlet JJs, in Fig.~\ref{fig:current}(a) and (b) show $I_S$ and $I_A$ at $T/T_c = 0.01$ and 0.5, respectively, where $T_c$ is the critical temperature of $S_L$. At the lower $T$ [Fig.~\ref{fig:current}(a)], the positive energy band is almost completely devoid of quasiparticle excitations, and tunneling between the $S_L$ and $S_R$ requires the traversal of the energy gap $\Delta_L + \Delta_R$. This can be achieved by the absorption of $n$ photons, leading to resonance peaks in the dissipationless $I_S$, as well as a step-like increase in the dissipative $I_A$ at $\hbar \Omega= (\Delta_L + \Delta_R)/n$. Here, $n = 1,2$, as limited by our truncation of the frequency modes. Including more modes would produce additional peaks, but these contributions are at most of the order $\alpha_1^4$ and are thus neglected. At the higher $T$ [Fig.~\ref{fig:current}(b)], thermally excited quasiparticles exist in the system, the positive energy band of $S_R$ is occupied and tunneling of these quasiparticles requires 
overcoming a gap of $\Delta_L - \Delta_R$. Hence, additional features emerge in both the 
dissipationless and dissipative currents at $\hbar\,\Omega =(\Delta_L - \Delta_R)/n$ that only appear at a sufficiently high $T$. 
 
Turning to the singlet-triplet JJs in Fig.~\ref{fig:current}(c) and (d), we find  
several key differences. The current is an order of magnitude larger because the effect is linear in $\alpha_1$, not quadratic, due to the selection rules for photon-assisted tunneling only allowing for odd powers in $\alpha_1$, which leads to fewer resonances. Indeed, at $T/T_c = 0.01$, in Fig.~\ref{fig:current}(c) we see that $I_S$ features a peak only at $\hbar\,\Omega \approx \Delta_L + \Delta_{0,R}$, 
which is also approximately at the onset of 
the dissipative $I_A$. There is a greater broadening than for singlet-singlet JJs because of the $p$-wave $\Delta_R$ is momentum dependent. At $T/T_c = 0.5$, in Fig.~\ref{fig:current}(d), we observe that the dissipationless $I_S$ is suppressed for $\hbar\,\Omega <\Delta_L - \Delta_R$. This is because the positive band edge in $S_L$ and $S_R$ are fully occupied by thermally excited quasiparticles at this $T$, effectively blocking the off-shell tunneling processes that carry the Cooper pair tunneling at subgap energies. The same is observed in the singlet-singlet JJs [Fig.~\ref{fig:current}(b)] at $\hbar\,\Omega < (\Delta_L - \Delta_R)/2 = 0.25\Delta$. 
In this regime, $I_A \neq 0$ because $\Delta_L > \Delta_R$, and the quasiparticles excited in $S_L$ easily tunnel to unoccupied states in $S_R$. 
For $\hbar\,\Omega >\Delta_L - \Delta_{R,0}$, both $I_S$ and $I_A$ are nonzero by the same mechanism as for the singlet-singlet  JJs. 

Typically, $V_G(t)$ in Fig.~\ref{fig:model} produces a slightly more complicated SOC that is better described by a Rashba model, $V(t) = \alpha(t)\left(k_x \sigma_y - k_y\sigma_x\right)$
to which we generalize our prior analysis. For a non-zero JE in singlet-triplet JJs, the requirement is that the triplet order parameter is aligned with the SOC in both spin and momentum space. This means that the $d$ vector must lie in the $xy$ plane. If we parameterize its spin and momentum structure as 
\begin{align}
\Delta_R = \Delta_{R,0}\left(\sigma_x\cos\xi + \sigma_y\sin\xi\right)\frac{k_x\cos\eta + k_y\sin\eta}{k_F},
\label{eq:sscurrent}
\end{align}
where $\xi$ ($\eta$) is the orientation in spin (momentum) space, 
we predict the angle-resolved current density $J(\theta,t) \propto
f(\theta,\xi,\eta) = \left(\cos^2\theta \sin\xi\cos\eta - \sin^2\theta \cos\xi\sin\eta\right)$. 
Considering also $d$-wave~\cite{SM}, our results suggest that by measuring the JE strength for different JJ orientations, the spin and momentum distribution of $\Delta_R$ could be inferred.

Our main focus in this work was to elucidate an unexplored dynamical Josephson effect with time-dependent SOC, including cases where the previous JJ studies used incompatible spin symmetries to exclude any Josephson current. A simple approach in Fig.~\ref{fig:phase}, even before our Green function calculations, already offers a deeper understanding of the selection rules, just as the optical selection rules have been recently revisited to include the role of topology and its implication on excitons~\cite{Zhang2018:PRL,Cao2018:PRL,Xu2020:PRL}. 

While generalizations of our model can be readily considered to make stronger connections with future experiments, we use the obtained results to estimate what could be expected in InAs-based 2DEG planar JJs~\cite{SM}, where the (quasistatic) $V_G$-controlled SOC was already demonstrated~\cite{Dartiailh2021:PRL,Mayer2020:NC,Zhou2022:NC}. Even though realizing the tens-of-GHz modulation is technically demanding, the availability of on-chip coplanar waveguides, impedance matching and gate stacks with low-capacitance~\cite{Tafuri:2019,Krantz2019:APR} make it experimentally feasible. Adjusting material transparency to reduce $\Delta_L$ and optimizing gate design could bring down the relevant frequencies to a more accessible $\sim10\;$GHz.

Our findings are not limited to describing previously excluded Josephson currents or offering another phase-sensitive probe of unconventional superconductors. Instead, they reveal that 
SOC-tunability of CPR can also be employed in novel qubits~\cite{Liu2025:X}, and to switch JJs~\cite{monroe2024}. Furthermore, it would be useful to consider other forms of SOC~\cite{Scharf2019:PRB}, including 
cubic in $k$, which transforms CPR as well as the symmetry of the superconducting correlations~\cite{Alidoust2021:PRB,amundsen2024b}. While we have focused on the tunneling limit, considering a higher junction transparency and SOC amplitude can enhance the predicted Josephson current and offer additional paths to design anharmonic CPR. 
There is also a much broader class of systems to test our predictions. Instead of the native triplet superconductivity, proximity-induced triplets can be realized with common $s$-wave superconductors and materials with the spin-dependent band structure due to static SOC in 2DEGs or structural chirality~\cite{Gorkov2001:PRL,amundsen2024b,Zhang2023:N,Xie2023:PRL}. Triplets can also be generated with ferromagnets, antiferromagets, and altermagnets~\cite{buzdin2005,Kontos2001:PRL,Banerjee2014:NC,Robinson2014,Singh2015:PRX,Bergeret2005:RMP,Jeon2023:NN,Ouassou2023:PRL,Zhu2025:X}, or by an applied Zeeman fields~\cite{Amet2016:S}.
Surprisingly, JE may even appear in junctions with a single superconductor~\cite{Gonzalez-Ruano2025:NC}. Common to such proximity systems is that singlet correlations will remain and contribute to the JE. However, this singlet-singlet JE with CPR $\propto \sin\phi$, which vanishes at $\phi = 0$, can be completely decoupled from our reported singlet-triplet JE, which attains its maximum at $\phi = 0$.

This work is supported by the Research Council of Norway
through its Centers of Excellence
funding scheme Grant No. 262633 “QuSpin”,
and by the U.S. Department
of Energy, Office of Science, Basic Energy Sciences under Award No. DE-SC0004890 (I.\v{Z}).

\bibliography{JJ_TDSOC}

\end{document}